# Ligand Pose Generation via QUBO-Based Hotspot Sampling and Geometric Triplet Matching

Pei-Kun Yang

E-mail: peikun@isu.edu.tw

ORCID: https://orcid.org/0000-0003-1840-6204

## Abstract

We propose a framework based on Quadratic Unconstrained Binary Optimization (QUBO) for generating plausible ligand binding poses within protein pockets, enabling efficient structure-based virtual screening. The method discretizes the binding site into a grid and solves a QUBO problem to select spatially distributed, energetically favorable grid points. Each ligand is represented by a three-atom geometric contour, which is aligned to the selected grid points through rigid-body transformation, producing from hundreds to hundreds of thousands of candidate poses. Using a benchmark of 169 protein–ligand complexes, we generated an average of 110 to 600,000 poses per ligand, depending on QUBO parameters and matching thresholds. Evaluation against crystallographic structures revealed that a larger number of candidates increases the likelihood of recovering near-native poses, with recovery rates reaching 100% for root mean square deviation (RMSD) values below 1.0 Å and 95.9% for RMSD values below 0.6 Å. Since the correct binding pose is not known in advance, we apply AutoDock-based scoring to select the most plausible candidates from the generated pool, achieving recovery rates of up to 82.8% for RMSD < 2.0 Å, 81.7% for RMSD < 1.5 Å, and 75.2% for RMSD < 1.0 Å. When poses with misleading scores are excluded, performance improves further, with recovery rates reaching up to 97.8% for RMSD < 2.0 Å and 1.5 Å, and 95.4% for RMSD < 1.0 Å. This modular and hardware-flexible framework offers a scalable solution for pre-filtering ligands and generating high-quality binding poses before affinity prediction, making it well-suited for large-scale virtual screening pipelines.

**Introduction**

Structure-based virtual screening (SBVS) requires access to the three-dimensional structures of both target proteins and candidate ligands (*1, 2*). When the structure of a protein–ligand complex is available (*3*), a variety of computational approaches can be used to estimate the binding free energy or binding score. These methods include detailed molecular dynamics simulations grounded in statistical thermodynamics (*4-6*), empirical scoring functions that simplify calculations to reduce computational complexity (*7-11*), and machine learning models trained on structure-based datasets (*12-14*). However, databases such as ZINC22 contain billions of small molecules (*15*), and experimental structures of their bound conformations are generally unavailable. Therefore, docking algorithms are commonly used to generate approximate binding geometries before any binding affinity prediction can be performed (*16-18*).

The success of SBVS largely depends on its ability to identify ligands with higher binding affinity. Regardless of the chosen screening method, the predicted affinity is highly sensitive to the accuracy of the ligand's binding pose. Most scoring functions are designed based on experimentally determined protein–ligand complexes, which means they are tuned to recognize near-native poses. In most SBVS pipelines, ligand poses are generated computationally using docking algorithms. Although the predicted poses may be within 2.0 Å root mean square deviation (RMSD) of the true conformation, such deviations can still cause significant errors in scoring functions, particularly in van der Waals repulsion. As a result, the scoring function must identify the most plausible pose from a set of candidates that may all deviate from the true binding mode, sometimes containing poses with unfavorable interactions such as high van der Waals repulsion.

Obtaining accurate predictions of binding affinity often involves balancing computational efficiency and precision. For example, the ZINC22 database includes more than 4.5 billion commercially available compounds with defined 3D structures (*15*). In large-scale screening tasks, it is necessary to exclude compounds that are unlikely to bind, to make the overall computation tractable. If too few binding poses are generated for each ligand, potentially active compounds may be missed simply because the correct binding geometry was not sampled. On the other hand, generating too many poses can lead to a significant increase in the cost of downstream filtering and analysis. A practical solution is to create a manageable number of high-quality poses. When these candidate poses are close to the true binding conformation, later stages, such as molecular dynamics-based free energy calculations, can converge faster, as less structural rearrangement is required.

From a thermodynamic perspective, both protein folding and ligand binding can be described as energy minimization problems. While such issues are easy to solve in low-dimensional cases using classical methods such as Newton–Raphson (*19*), finding global minima in high-dimensional energy landscapes remains extremely challenging. To address this, specialized hardware platforms like quantum annealers from D-Wave (*20*) and digital annealers developed by Fujitsu (*21*) have been introduced to solve quadratic unconstrained binary optimization (QUBO) problems efficiently. Compared to conventional combinatorial optimization techniques (*22-25*), these novel platforms offer faster and more scalable solutions, especially for problems involving large numbers of binary variables (*26*).

The QUBO problem can be formulated as a quadratic function of binary variables, where the objective is to find the binary vector *X* that minimizes the function *F*(*X*), defined as:

$$F(X) = X^\mathrm{T} J X + H X \tag{1}$$

Given matrices **J** and **H**, the goal is to find *X* that minimizes *F*(*X*). In this study, we construct a QUBO model to identify ligand binding poses by discretizing the binding pocket into a 3D grid. The solver then determines which grid points are most likely to be occupied by ligand atoms. By aligning these grid-based selections with ligand atomic geometries, we generate plausible binding poses. A PyTorch-based QUBO solver is employed, enabling efficient execution on widely available GPU hardware (*27*).

**Methods**

**Ligand Placement via QUBO-Based Hotspot Optimization.** To identify favorable spatial arrangements of ligand atoms within a protein's binding pocket, we developed a stepwise computational workflow that proceeds through three major stages. First, we identify grid points within the pocket that exhibit strong energetic attraction using a QUBO-based selection strategy. Next, we guide the positioning of the ligand by applying geometric information derived from its structure. Finally, we evaluate the resulting binding poses to determine their plausibility and quality.

The binding site of the target protein is first discretized into a three-dimensional grid. Each grid point is assigned a numerical value that reflects its van der Waals interaction energy with nearby protein atoms. These values serve as input to construct a QUBO formulation, which aims to select a subset of grid points that are not only energetically favorable but also spatially distributed throughout the pocket. This selected subset defines potential locations for ligand atoms.

To guide the placement of ligands, we extract a geometric contour from each molecule by selecting three atoms that define its orientation in space. These three atoms serve as a reference triangle for spatial alignment. The ligand is then rigidly aligned to a selected triplet of QUBO-chosen grid points that best match this geometric contour, producing candidate poses based on this mapping.

After the ligand has been positioned in the pocket, each candidate pose is evaluated using empirical scoring functions. The pose that receives the most favorable score, typically associated with the lowest predicted binding free energy, is selected for output. This structure can then be used for further analysis, including binding affinity estimation or pose refinement.

This overall approach allows us to efficiently generate diverse and physically meaningful ligand poses by balancing both energetic preferences and geometric compatibility. The use of a PyTorch-based QUBO solver further accelerates computation by leveraging GPU resources.

An overview of the entire ligand placement workflow, including QUBO-based hotspot detection, geometric alignment, and scoring, is illustrated in Figure 1.

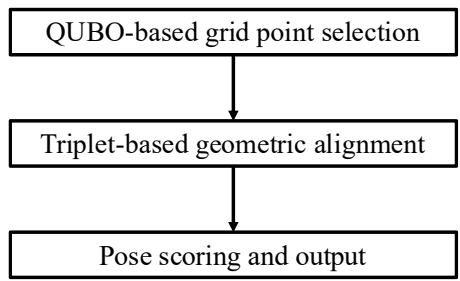

**Figure 1. Overview of the QUBO-based ligand placement workflow.** The process begins with identifying energetically favorable and spatially distributed grid points using QUBO optimization. A geometric contour of the ligand, defined by three representative atoms, is aligned to these points to generate candidate poses. Each pose is then evaluated using scoring functions to select the final predicted binding conformation.

**QUBO-Based Grid Encoding for Ligand Placement.** To model the spatial and energetic characteristics of ligand binding, the binding pocket is represented as a spherical grid centered at the origin, with a radius of 7 Å and a grid spacing of 0.375 Å. This setup produces a total of 27,201 discrete grid points. The coordinates of the protein and ligand are translated so that the centroid of the ligand is aligned to the origin, thereby preserving their spatial relationship in the grid-based representation.

Each grid point corresponds to a binary variable that indicates whether it is selected as part of a predicted ligand pose. The collection of these binary variables forms the input vector for the QUBO model. The QUBO objective function is defined by two primary components: a pairwise matrix that enforces geometric constraints, and a unary energy vector that captures the strengths of protein-ligand interactions.

The pairwise interaction matrix accounts for distances between all grid point pairs. If two selected points are too close, they are penalized to prevent steric clashes. If their distance falls within a defined optimal range, they contribute positively to the QUBO score, promoting spatial regularity. Pairs that fall outside the preferred range do not influence the objective.

To formally define this preferred distance window, we introduce a central value $\bar{d}$ and a tolerance $\delta$, resulting in the interval $[\bar{d} - \delta, \bar{d} + \delta)$. Grid point pairs with distances within this interval are rewarded with negative energy values, encouraging their selection during optimization. Pairs that are too close are strongly penalized to avoid steric clashes, while those outside the interval do not contribute to the objective function. This formulation improves placement quality and will be referenced again in the Results section when evaluating spatial distribution and coverage.

The energy vector represents van der Waals interaction energies between a hypothetical carbon atom placed at each grid point and the surrounding protein atoms. These interaction energies are obtained from precomputed potential maps using AutoGrid4 (*17*). Negative energy values signify attraction and are rewarded in the QUBO model.

By minimizing the QUBO objective, the solver identifies a subset of grid points that together offer both favorable interaction energies and spatial feasibility. The optimization is performed using a differentiable PyTorch-based solver, which relaxes the binary variables to continuous values using sigmoid activation and then applies a threshold to generate the final binary selection (*27*).

**Dataset Preparation.** Protein–ligand complexes were obtained from the CASF-2016 benchmark set, which contains 285 high-quality crystal structures of protein–ligand pairs. For each complex, van der Waals energy maps were generated using AutoGrid4, a component of the AutoDock4 suite. During preprocessing, we excluded 12 complexes for which AutoGrid4 lacked built-in parameter templates for certain atom types, resulting in 273 structures with successfully generated energy maps.

To ensure meaningful interaction with the binding site, we further filtered the dataset based on spatial proximity. Specifically, only ligands whose heavy atoms could be fully enclosed within a 7 Å radius centered at the designated binding pocket were retained. After this spatial filtering step, a total of 169 protein–ligand complexes remained and were used for downstream QUBO formulation and analysis. All protein and ligand structures were converted to PDBQT format for compatibility with AutoDock tools.

**Ligand Contour Definition via Three-Atom Selection.** To define the spatial contour of the ligand, we begin by identifying three representative non-hydrogen atoms. All pairwise distances between heavy atoms are computed, and the atom pair $(i, j)$ with the largest distance $d_{ij}$ is selected to define the principal axis of the ligand. This step also enables rapid filtering based on molecular size. A third atom $k$ is then selected such that the sum $d_{ik} + d_{jk}$ is maximized. This selection strategy ensures that atoms $i, j, k$ capture the ligand's overall spatial extent and orientation while avoiding collinearity.

**Geometric Matching to Grid Triplets.** For each selected ligand triangle $(i, j, k)$, we identify candidate grid triplets $(i', j', k')$ among those selected by the QUBO solver by comparing all three pairwise distances. A grid triplet is retained only if the corresponding distances $d_{i'j'}$, $d_{i'k'}$ and $d_{j'k'}$ each fall within the predefined interval $[\bar{d} - \delta, \bar{d} + \delta)$, matching the ligand's $d_{ij}$, $d_{ik}$ and $d_{jk}$, respectively. If any one of the distances falls outside this interval, the triplet is rejected.

Once a valid triplet is identified, a rigid-body transformation is computed that aligns the ligand atoms ($i$, $j$, $k$) to the matched grid points ($i'$, $j'$, $k'$). This transformation is then applied to the entire ligand structure to generate a candidate binding pose.

**Scoring and Final Pose Selection.** To rank candidate poses, we employed the AutoDock 4.2 scoring function to estimate the ligand's binding free energy at each candidate location. The pose with the lowest predicted binding free energy was selected as the final prediction. The RMSD between this pose and the experimental structure was then calculated to evaluate accuracy.

**Results**

**How Distance Parameters Influence Grid Point Selection in QUBO.** The way our QUBO model selects grid points is shaped primarily by two key parameters. One is the average distance between selected points, which we call $\bar{d}$, and the other is the allowed tolerance around that average, denoted as $\delta$.

When $\bar{d} \pm \delta = 1.0 \pm 0.3$ Å, the model identifies the highest number of valid points, reaching about 2,050. However, increasing the average distance to 1.5 Å, even with wide tolerances such as ±0.5 Å or ±0.4 Å, significantly reduces the number of selected points, down to 918 and 597, respectively. This indicates that as the spacing requirement increases, fewer point combinations satisfy the condition.

On the other hand, keeping $\bar{d} = 1.0$ Å but narrowing the tolerance from ±0.3 to ±0.2 also reduces the point count, from 2,050 to 1,066. These observations demonstrate that both parameters directly influence the number and spatial distribution of selected points. The average distance determines how far apart points should be, while the tolerance controls how strictly that rule is applied.

As illustrated in Figure 2, different parameter settings yield visibly distinct point distributions across the binding pocket of protein 1A30.

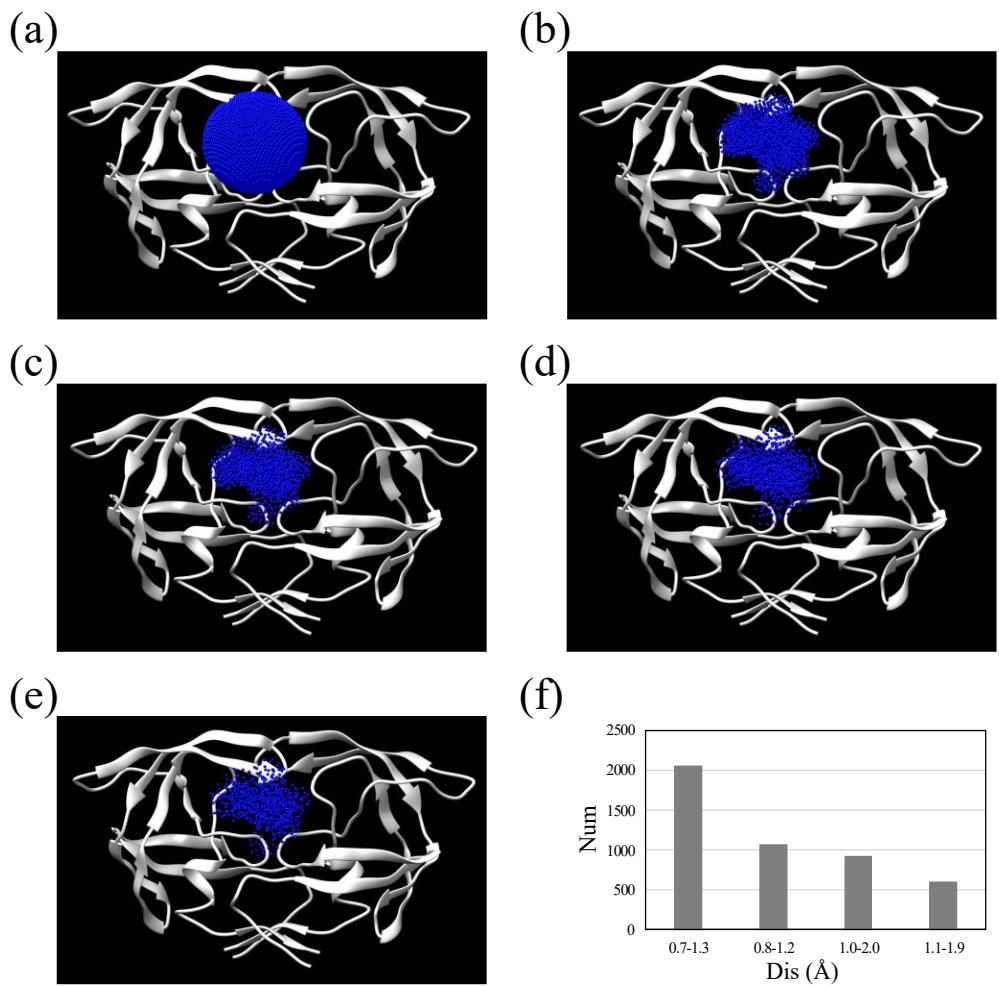

**Figure 2. Effect of Distance Constraints on QUBO-Based Grid Point Selection in Protein 1A30.** Panel (a) displays the structure of protein 1A30 (PDB ID: 1A30) overlaid with a spherical grid (radius = 7 Å, spacing = 0.375 Å). Panels (b–e) show grid points selected by the QUBO model under different average distance constraints ($\bar{d} \pm \delta$) encoded in the $\boldsymbol{J}$ matrix: (b) 1.0 ± 0.3 Å, (c) 1.0 ± 0.2 Å, (d) 1.5 ± 0.5 Å, and (e) 1.5 ± 0.4 Å. Panel (f) quantifies the number of selected points across each setting, illustrating that smaller $\bar{d}$ values and looser tolerances ($\delta$) lead to denser and more numerous selections, whereas tighter constraints reduce point count and spatial coverage.

**Representing Ligand Geometry with Three Atoms.** To assess whether a ligand can fit into a protein binding pocket, we first need a way to describe its spatial geometry. We do this by selecting three heavy atoms from the ligand to form a triangle that defines its shape and orientation in three-dimensional space.

We begin by calculating all pairwise distances between heavy atoms in the ligand. The pair with the largest distance defines the principal axis of the ligand, and a third atom is selected to maximize the total distance to the first two. This ensures that the selected atoms are spread apart, avoiding collinearity and forming a well-defined triangle.

This method provides a compact and robust geometric representation that accurately captures the overall shape of the ligand. As shown in Figure 3, the selected atoms are highlighted for several example ligands with different structures and flexibilities.

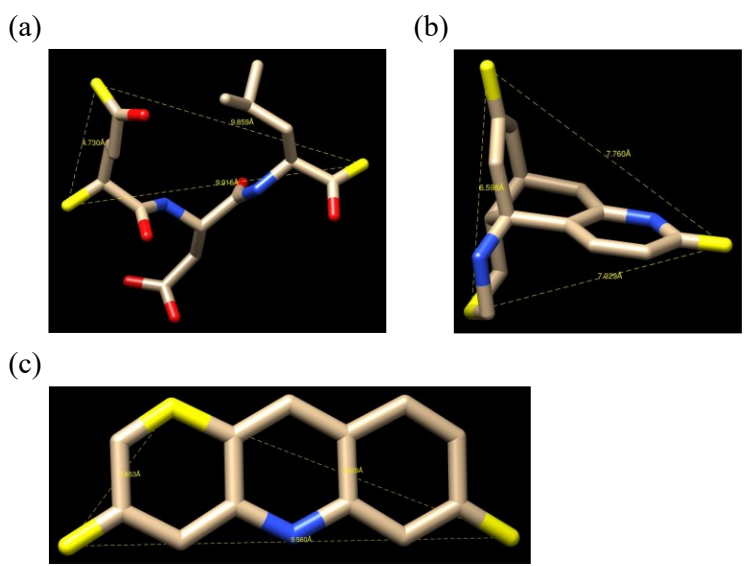

**Figure 3. Geometric definition of ligand contour using three representative atoms.** Three atoms were selected from each ligand as yellow spheres to define its spatial orientation and overall geometry. The selected atoms are connected by dashed lines indicating interatomic distances (in Å). (a) Ligand TRIPEPTIDE GLU-ASP-LEU (PDB ID: 1A30); (b) Ligand HUPERZINE B (PDB ID: 1GPN); (c) Ligand PROFLAVIN (PDB ID: 1BCU).

**Matching Ligand Contours to Grid Triplets.** Once we have a triangle defined by three ligand atoms, we search for matching triplets among the grid points selected by the QUBO model. Each candidate triplet is evaluated by comparing its three inter-point distances to those of the ligand triangle. A match is accepted only if all three distances fall within a predefined tolerance.

When a match is found, we compute a rigid-body transformation that aligns the ligand triangle to the grid triplet. This transformation is then applied to the entire ligand to generate a candidate pose. As shown in Figure 4a, this approach enables us to position the ligand within the pocket in a manner that respects both its shape and the spatial arrangement of favorable grid positions. Figure 4b illustrates how varying the matching threshold impacts the number of accepted placements.

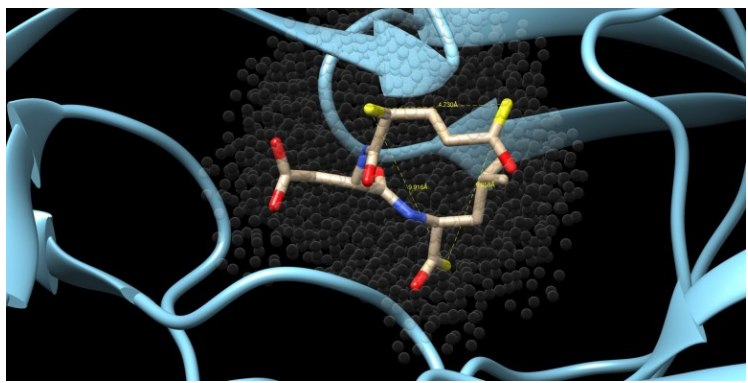

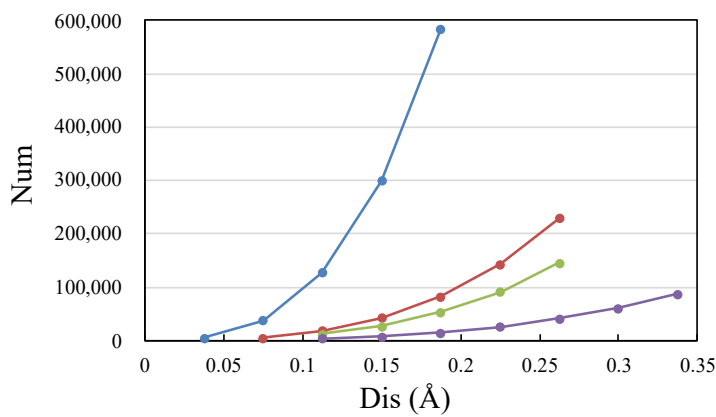

**Figure 4. Ligand placement through three-atom contour matching in a QUBO-derived grid.** (a) Example of ligand placement using PDB ID: 1A30. The ligand is positioned by aligning its geometric contour, defined by three non-hydrogen atoms (yellow), to QUBO-selected grid points (black spheres), which represent spatially favorable and energetically attractive locations within the binding pocket. (b) Effect of matching threshold on the number of accepted placements. The y-axis shows the average number of retained poses per ligand across the dataset, and the x-axis indicates the maximum allowed deviation in pairwise distances between ligand atoms and matched grid points. Stricter thresholds result in fewer valid alignments. Colors indicate different RMSD windows used during matching: blue corresponds to $1.0 \pm 0.3$ Å, red to $1.0 \pm 0.2$ Å, green to $1.5 \pm 0.5$ Å, and purple to $1.5 \pm 0.4$ Å.

**How Closely Do the QUBO-Based Ligand Placements Match the Experimental Structures?** To evaluate how well our predicted ligand poses align with experimentally determined structures, we used the RMSD between each predicted pose and its corresponding crystal structure, based on all heavy atoms.

As shown in Figure 5a, the predicted pose for HIV protease (PDB ID: 1A30) aligns closely with the crystal ligand. We applied the same comparison across 169 protein–ligand complexes and analyzed recovery rates under different RMSD thresholds, specifically at 2.0, 1.5, 1.0, and 0.6 Å.

Figures 5b through 5e reveal a clear trend. When the RMSD threshold is relatively loose, such as 2.0 or 1.5 Å, the framework consistently achieves high recovery rates, even with as few as 5,000 candidate poses per ligand. Specifically, at RMSD < 2.0 Å, a pose count of around 5,000 is sufficient to reach 100% recovery. At RMSD < 1.5 Å, the same pose count still yields approximately 99% accuracy. As the threshold becomes stricter, such as 1.0 Å or 0.6 Å, more candidate poses are required to maintain high accuracy. For RMSD < 1.0 Å, approximately 20,000 poses per ligand can achieve up to 98% recovery, while reaching 95% recovery at RMSD < 0.6 Å typically requires a similar number of poses. These observations highlight the importance of candidate pose count when aiming for sub-angstrom precision.

Importantly, our results suggest that these differences are less about the specific QUBO parameters themselves, such as the chosen values of $\bar{d}$ and $\delta$, and more about how many ligand poses were generated under each configuration. Settings that produced more candidates tended to achieve higher recovery, especially under tight RMSD thresholds. In other words, the placement accuracy is more closely linked to the size of the candidate pool than to the exact spacing constraints used during grid selection.

This finding highlights the practical importance of pose sampling density in ligand docking. As long as a diverse and sufficiently large set of candidate poses is generated, the QUBO-based method consistently includes near-native conformations among the predicted options.

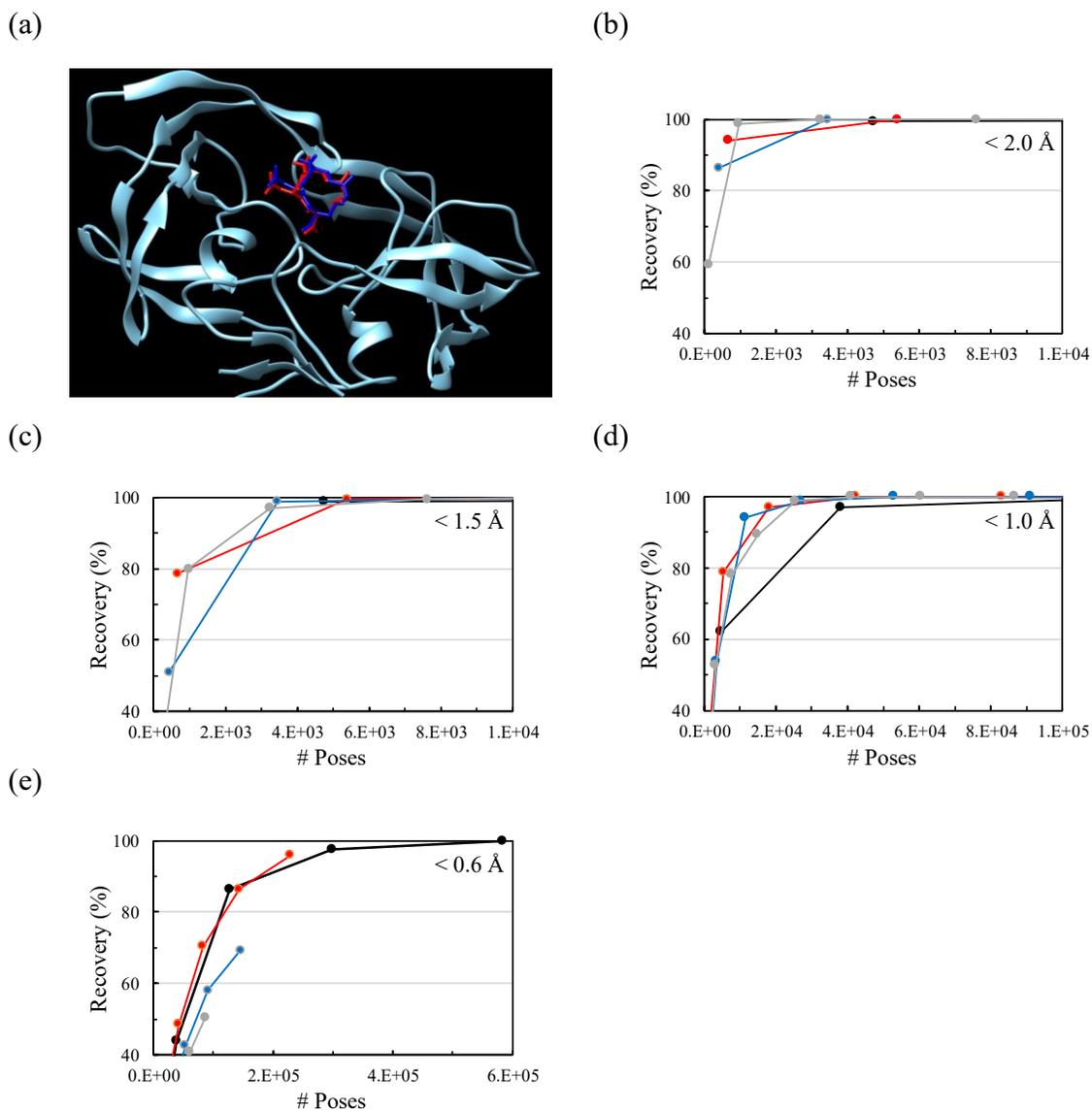

**Figure 5. Accuracy of ligand placements generated by QUBO-based geometric alignment.** (a) Experimental ligand pose (red) and its closest QUBO-predicted counterpart (blue) in HIV protease (PDB ID: 1A30), aligned via three selected atoms. (b–e) Recovery rates of predicted poses within RMSD thresholds of 2.0, 1.5, 1.0, and 0.6 Å, based on all non-hydrogen atoms. The x-axis shows the number of candidate poses, while curves correspond to different $J$ matrix settings: $1.0 \pm 0.3$ Å (black), $1.0 \pm 0.2$ Å (red), $1.5 \pm 0.5$ Å (blue), and $1.5 \pm 0.4$ Å (gray).

**Picking the Best Pose Using AutoDock Scoring.** After generating a diverse set of poses, we need a way to choose the best one. In this study, we used the AutoDock scoring function to estimate the binding free energy of each pose. As shown in Figure 6 and further detailed in Table A.I, this approach allowed us to evaluate the impact of different geometric matching thresholds on pose quality. The modularity of our pipeline means that this step can be easily replaced by other scoring strategies, such as deep learning models or knowledge-based functions, depending on the application.

Figures 6a through 6c reveal a consistent trend. At RMSD < 2.0 Å, generating approximately 10,000 candidate poses per ligand is sufficient to achieve a recovery rate of around 80%. At RMSD < 1.5 Å, a similar number of poses still yields approximately 78% recovery. As the threshold becomes stricter, such as RMSD < 1.0 Å, a significantly larger number of candidate poses, approximately 220,000, is required to achieve 75% recovery. These results highlight the trade-off between alignment precision and the number of required poses. Table A.I further shows that some predicted poses achieve a lower binding free energy than the experimentally determined structure, yet are located farther from the native binding site.

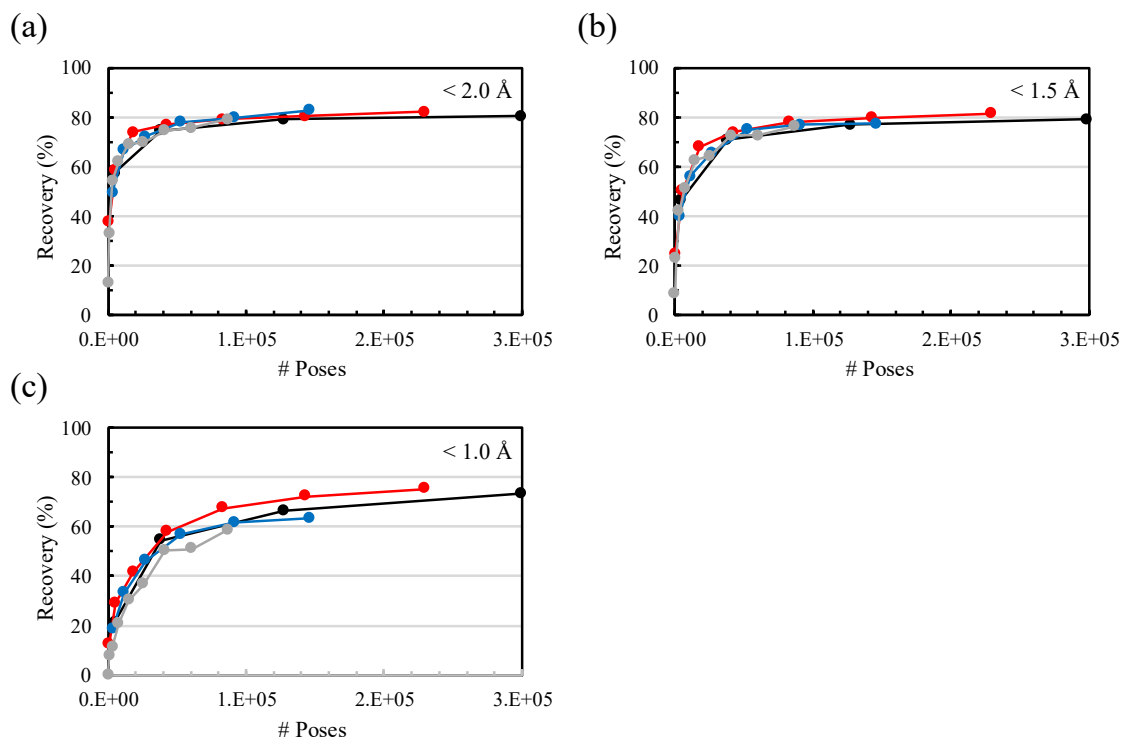

**Figure 6. Accuracy of AutoDock selected ligand poses under varying RMSD thresholds.** Recovery rates of poses selected by AutoDock based on predicted binding free energy. The y-axis shows the percentage of poses within each RMSD cutoff: (a) 2.0 Å, (b) 1.5 Å, and (c) 1.0 Å, compared to the crystal structure. The x-axis indicates the number of QUBO generated candidate pose. Curves correspond to QUBO settings: 1.0 ± 0.3 Å (black), 1.0 ± 0.2 Å (red), 1.5 ± 0.5 Å (blue), and 1.5 ± 0.4 Å (gray).

**Improving Performance by Filtering Misleading Scoring Results.** AutoDock performs reasonably well in most cases, but it is not perfect. As illustrated in Figure 7 and supported by the quantitative comparisons in Table A.I, there were situations where poses with high RMSD values were scored more favorably than the near-native structures. These mis-rankings are problematic because they may result in selecting incorrect poses during virtual screening.

To address this, we applied a filtering step based on the analysis shown in Table A.1. For each RMSD threshold (2.0, 1.5, and 1.0 Å), we identified poses that exceeded the threshold yet received a lower binding free energy score than the experimental pose. These misleading poses were excluded from the candidate set. After removing such energetically favorable but geometrically incorrect poses, the accuracy of selecting near-native conformations improved considerably.

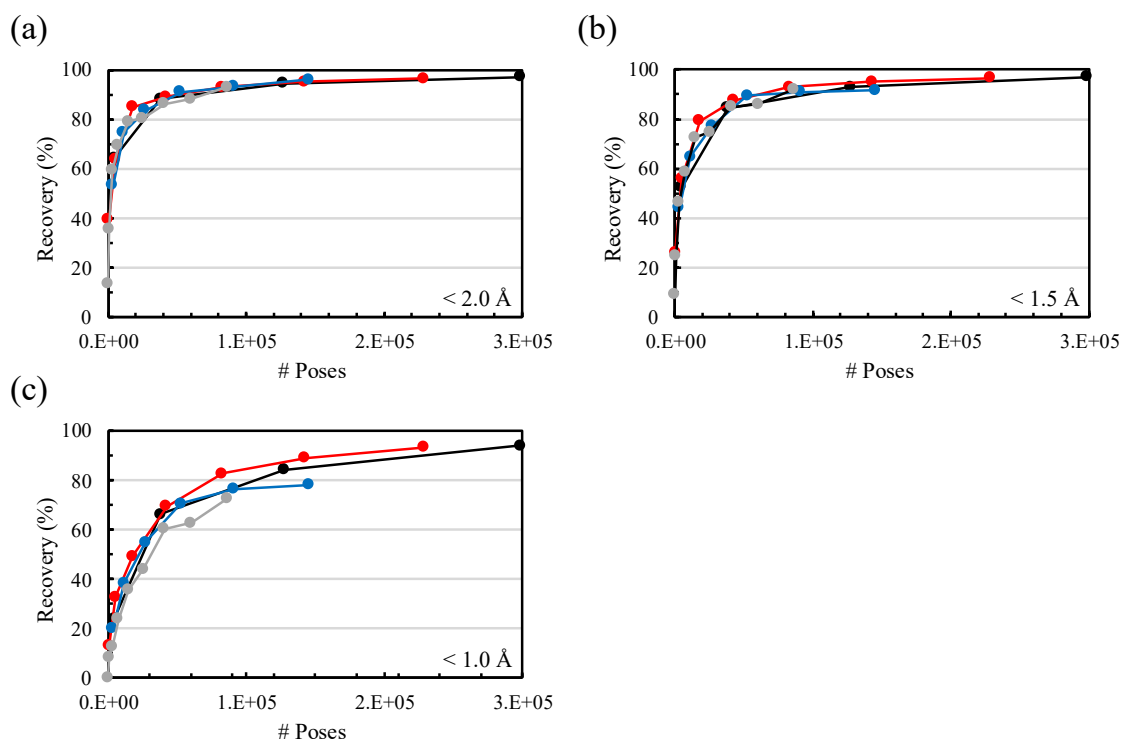

**Figure 7. Improved pose recovery after excluding scoring outliers.** Ligand poses with high RMSD were filtered out, as favorable AutoDock scores were used to reduce misranked predictions. Shown are updated recovery rates for RMSD thresholds of (a) 2.0 Å, (b) 1.5 Å, and (c) 1.0 Å. Curves represent QUBO settings as in Figure 6.

**Discussions**

**QUBO-Based Sampling Strategy.** To figure out where a ligand might fit inside a protein binding pocket, we started by dividing the space into a spherical 3D grid. We chose a radius of 7 Å and a spacing of 0.375 Å, resulting in 27,201 grid points. That number is large enough to cover the pocket comprehensively, yet still small enough to keep the problem manageable. If we had made the grid spacing smaller, we could have achieved higher spatial precision, but the number of variables would have increased dramatically. So we settled on these values as a compromise between resolution and computational cost.

Our QUBO formulation relies on two essential components: the ***H*** matrix and the ***J*** matrix. The ***H*** matrix reflects van der Waals interactions between each grid point and the protein. Grid points that show attractive interactions are assigned negative energy values, making them more favorable for selection. On the other hand, grid points that would result in repulsion are penalized, making them less likely to be chosen.

The ***J*** matrix helps control the spacing of the selected points. It does this by evaluating every pair of grid points and encouraging those whose distances fall within a defined range. This range is specified using a target average distance, denoted as $\bar{d}$, along with a tolerance $\delta$. If $\bar{d}$ is small, the selected points tend to be more densely packed. If $\delta$ is larger, there is more flexibility in how far apart the points can be. By adjusting these two parameters, users can control the number of points selected and their distribution, depending on the ligand's size and shape. These effects are visible in Figure 2.

**Solver Considerations and Implementation.** While dedicated hardware, such as quantum annealers or digital annealers, is designed for solving QUBO problems, these platforms are often expensive or difficult to access. In this project, we instead used a PyTorch-based simulator to solve QUBO problems on regular CPUs and GPUs. PyTorch offered two major advantages. First, it is free and open-source. Second, it supports GPU acceleration with minimal setup, making it easy to scale up our computations for large-scale screening.

Of course, QUBO solvers, whether quantum or classical, often yield different results in separate runs due to the inherent randomness in the optimization process. However, by analyzing a large number of protein–ligand systems, we were able to shift our focus from individual outputs to broader patterns. This allowed us to draw consistent conclusions about spatial sampling, even without relying on specialized hardware. With careful setup and thoughtful use of GPU resources, we found that QUBO-based modeling remains both accessible and effective when using commonly available computational tools.

**Ligand Contour Definition Using Three Reference Atoms.** Before deciding whether a ligand fits into the pocket, we need to understand how much space the ligand occupies. We start by identifying the two atoms in the ligand that are farthest apart. This gives us a rough idea of the ligand's maximum extent. We then examine the QUBO-selected grid points to identify pairs that match this distance within a specified range. This step ensures that the ligand's longest internal dimension can be placed inside the pocket.

But two atoms are not enough to fix the ligand's full 3D orientation. To resolve this, we add a third atom. Specifically, we look for the atom that is farthest from both of the original two. More precisely, we select the one that maximizes the sum of distances to the first pair. This gives us a triangle that defines both the size and the spatial orientation of the ligand. It might not always be the mathematically perfect triangle, but it works well in practice and is simple to compute.

Using three atoms makes it possible to align the ligand to a set of grid points using rigid-body transformations. If we had used more atoms, we could reduce ambiguity even further; however, this would also make the geometric comparisons much more computationally expensive. For large-scale screening, using three atoms gives us a good balance between accuracy and speed.

**Triplet Matching Based on Ligand Contour.** Once we find a grid point pair that matches the ligand's longest internal distance, the next step is to locate a third grid point that completes the triangle. This third point needs to have the right distances to the first two, within a tolerance range. We only keep triplets where all three distances match their counterparts in the ligand triangle.

If we use a large tolerance, we end up accepting more triplets. This increases the number of candidate poses, which can improve our chances of finding a good match. But it also means we have to evaluate more poses, which increases computational cost. So the choice of matching tolerance reflects a trade-off between accuracy and efficiency. The effects of this trade-off can be seen in Figure 4b.

**Impact of Triplet Quantity on Placement Accuracy.** Once we have a matched triplet, we perform a rigid-body transformation to align the ligand's triangle onto the grid points. This transformation is applied to the entire ligand to generate a candidate binding pose. We then compare this pose to the known crystal structure using RMSD calculated over all heavy atoms.

There are two main ways to increase the number of candidate poses. One is to adjust the QUBO parameters, particularly the values of $\bar{d}$ and $\delta$ in the $J$ matrix. A slightly looser definition of what constitutes a good pair of grid points will allow more combinations to qualify. The second is to increase the geometric matching threshold, which allows more grid triplets to match the ligand triangle.

As Figure 5 shows, increasing the number of candidate poses leads to better performance, especially when using stricter RMSD thresholds like 1.0 Å or 0.6 Å. This pattern holds regardless of whether the additional diversity originates from QUBO settings or the matching criteria. If computational resources permit, generating more poses is a reliable way to enhance the likelihood of recovering native-like structures.

**Influence of Scoring Function on Pose Selection.** After generating a diverse set of poses, we need a way to choose the best one. In our study, we used the AutoDock scoring function to estimate binding free energy. But this framework is modular, which means we could just as easily use another scoring method, such as a machine learning model or a knowledge-based function.

AutoDock performs reasonably well in most cases, but it is not perfect. There were instances where the pose with the best AutoDock score was not the most accurate one in terms of RMSD. This is a known limitation of many scoring functions.

To address this, we filtered out cases where a high RMSD pose had a better score than a near-native pose. After applying this filter, the overall accuracy of pose selection improved noticeably. The improvements are shown clearly in Figure 7. This simple filtering step demonstrates that scoring functions are helpful but not flawless.

**Implications for Virtual Screening Pipelines.** This QUBO-based method introduces a novel approach to virtual screening. One of the key features of our approach is that we break the task into two stages. First, we use the QUBO solver to pick out promising grid points. Second, we match ligand substructures to those points based on geometry.

This design has several advantages. The grid point selection does not depend on the ligand, which means we can compute it once and use it for any number of ligands targeting the same protein. The geometric matching step also acts as a filter. If a ligand is too large to fit in the pocket, we can exclude it early on, which saves time in downstream scoring.

The system is also flexible in terms of the tools you use. In our case, we used PyTorch and AutoDock, but the same pipeline could be extended with different scoring functions or accelerated solvers. Whether you are using CPUs, GPUs, or quantum hardware, this approach can be adapted to your specific setup.

Overall, this framework is scalable, modular, and practical. It enhances the efficiency and adaptability of structure-based virtual screening, particularly when working with large ligand libraries or challenging protein targets.

**Conclusion**

We developed a modular and efficient method for generating ligand poses by integrating QUBO-based grid sampling with triplet-based geometric matching. By identifying favorable grid points in advance, we reduce the need to explore all possible placements through exhaustive search. Representing each ligand with three reference atoms enables fast geometric alignment and helps preserve structural integrity. This two-step process, which separates spatial sampling from geometric alignment, provides an effective way to eliminate ligands that are too large or incompatible with the binding pocket. Importantly, our results show that the ability to recover the correct binding pose depends on the number of candidate poses generated, which can be adjusted through QUBO parameters and geometric thresholds. This allows users to balance accuracy and computational cost based on their target precision and available resources. Although we employed the AutoDock scoring function in this study, the framework is modular and can be extended to support alternative scoring strategies, including machine learning-based or knowledge-based approaches. Overall, this scalable and flexible design makes our method well suited for large-scale virtual screening pipelines where efficiency, adaptability, and pose quality are essential.

**Competing interests**

The authors declare no competing interests.

**Data and Software Availability**

The data supporting the findings of this study are openly available on GitHub at the following URL: https://github.com/peikunyang/02_QUBO.

**Reference**

1. E. H. B. Maia, L. C. Assis, T. A. De Oliveira, A. M. Da Silva, A. G. Taranto, Structure-based virtual screening: from classical to artificial intelligence. *Frontiers in chemistry* **8**, 343 (2020).

2. A. Varela-Rial, M. Majewski, G. De Fabritiis, Structure based virtual screening: Fast and slow. *Wiley Interdisciplinary Reviews: Computational Molecular Science* **12**, e1544 (2022).

3. S. K. Burley *et al.*, RCSB Protein Data Bank (RCSB. org): delivery of experimentally-determined PDB structures alongside one million computed structure models of proteins from artificial intelligence/machine learning. *Nucleic Acids Research* **51**, D488-D508 (2023).

4. H. Fu *et al.*, Accurate determination of protein: ligand standard binding free energies from molecular dynamics simulations. *Nature Protocols* **17**, 1114-1141 (2022).

5. G. Heinzelmann, M. K. Gilson, Automation of absolute protein-ligand binding free energy calculations for docking refinement and compound evaluation. *Scientific reports* **11**, 1-18 (2021).

6. E. Wang *et al.*, VAD-MM/GBSA: A Variable Atomic Dielectric MM/GBSA Model for Improved Accuracy in Protein–Ligand Binding Free Energy

Calculations. *Journal of Chemical Information and Modeling*,  (2021).

7. J. Dittrich, D. Schmidt, C. Pfleger, H. Gohlke, Converging a knowledge-based scoring function: DrugScore2018. *Journal of chemical information and modeling* **59**, 509-521 (2018).

8. J. Bao, X. He, J. Z. Zhang, Development of a New Scoring Function for Virtual Screening: APBScore. *Journal of Chemical Information and Modeling* **60**, 6355-6365 (2020).

9. C. N. Cavasotto, M. G. Aucar, High-throughput docking using quantum mechanical scoring. *Frontiers in chemistry* **8**, 246 (2020).

10. M. Kadukova, K. d. S. Machado, P. Chacón, S. Grudinin, KORP-PL: a coarse-grained knowledge-based scoring function for protein–ligand interactions. *Bioinformatics* **37**, 943-950 (2021).

11. C. Yang, Y. Zhang, Lin_F9: A Linear Empirical Scoring Function for Protein–Ligand Docking. *Journal of Chemical Information and Modeling*,  (2021).

12. C. Schneider, A. Buchanan, B. Taddese, C. M. Deane, DLAB: deep learning methods for structure-based virtual screening of antibodies. *Bioinformatics* **38**, 377-383 (2022).

13. H. Li, K. H. Sze, G. Lu, P. J. Ballester, Machine-learning scoring functions for structure-based virtual screening. *Wiley Interdisciplinary Reviews: Computational Molecular Science* **11**, e1478 (2021).

14. J. Ricci-Lopez, S. A. Aguila, M. K. Gilson, C. A. Brizuela, Improving structure-based virtual screening with ensemble docking and machine learning. *Journal of Chemical Information and Modeling* **61**, 5362-5376 (2021).

15. B. Tingle *et al.*, ZINC-22-A free multi-billion-scale database of tangible compounds for ligand discovery.  (2022).

16. A. T. McNutt *et al.*, GNINA 1.0: molecular docking with deep learning. *Journal of cheminformatics* **13**, 1-20 (2021).

17. J. Eberhardt, D. Santos-Martins, A. F. Tillack, S. Forli, AutoDock Vina 1.2. 0: New docking methods, expanded force field, and python bindings. *Journal of Chemical Information and Modeling* **61**, 3891-3898 (2021).

18. W. J. Allen *et al.*, DOCK 6: Impact of new features and current docking performance. *Journal of computational chemistry* **36**, 1132-1156 (2015).

19. T. J. Ypma, Historical development of the Newton–Raphson method. *SIAM review*

**37**, 531-551 (1995).

20. D. Willsch *et al.*, Benchmarking Advantage and D-Wave 2000Q quantum annealers with exact cover problems. *Quantum Information Processing* **21**, 141 (2022).

21. O. Şeker, N. Tanoumand, M. Bodur, Digital annealer for quadratic unconstrained binary optimization: a comparative performance analysis. *Applied Soft Computing* **127**, 109367 (2022).

22. B. D. Woods, G. Kochenberger, A. P. Punnen, in *The Quadratic Unconstrained Binary Optimization Problem*. (Springer, 2022), pp. 301-311.

23. M. Zaman, K. Tanahashi, S. Tanaka, PyQUBO: Python library for mapping combinatorial optimization problems to QUBO form. *IEEE Transactions on Computers* **71**, 838-850 (2021).

24. G. Kochenberger *et al.*, The unconstrained binary quadratic programming problem: a survey. *Journal of combinatorial optimization* **28**, 58-81 (2014).

25. G. Tavares, *New algorithms for Quadratic Unconstrained Binary Optimization (QUBO) with applications in engineering and social sciences*. (Rutgers The State University of New Jersey-New Brunswick, 2008).

26. Y. W. Koh, H. Nishimori, Quantum and classical annealing in a continuous space with multiple local minima. *Physical Review A* **105**, 062435 (2022).

27. P.-K. Yang, Comparative Evaluation of PyTorch, JAX, SciPy, and Neal for Solving QUBO Problems at Scale. *arXiv:2507.17770*, (2025).

Appendix

Table A.I. AutoDock binding scores and RMSD values under different geometric matching thresholds. The matching threshold refers to the maximum allowed deviation in pairwise distances (in Å) between ligand atoms and QUBO-selected grid points, with $\bar{d} \pm \delta = 1.0 \pm 0.3$ Å. For each threshold (0.5, 0.4, 0.3, 0.2, 0.1 Å), the best-scoring pose was selected and its predicted binding energy (in kcal/mol) and RMSD (in Å) to the experimental ligand pose are shown. The "Exp" column reports the AutoDock score computed using the crystallographic ligand position.

|  | Exp | 0.5 | | 0.4 | | 0.3 | | 0.2 | | 0.1 | |
|---|---|---|---|---|---|---|---|---|---|---|---|
|  | score | score | rmsd | score | rmsd | score | rmsd | score | rmsd | score | rmsd |
| 1a30 | -5.37 | -6.09 | 6.04 | -6.02 | 6.16 | -6.02 | 6.16 | -5.68 | 6.02 | -4.83 | 7.12 |
| 1bcu | -4.73 | -5.65 | 6.99 | -5.65 | 6.99 | -5.65 | 6.99 | -5.07 | 2.63 | -4.76 | 3.4 |
| 1c5z | -6.51 | -6.02 | 2.47 | -5.92 | 0.6 | -5.91 | 2.47 | -5.91 | 2.47 | -5.33 | 2.54 |
| 1gpn | -8.37 | -8.98 | 4.28 | -8.98 | 4.73 | -8.95 | 4.75 | -8.3 | 4.32 | -8.18 | 4.41 |
| 1nc1 | -10.35 | -9.25 | 0.48 | -8.89 | 0.58 | -8.71 | 0.76 | -5.3 | 0.69 | -2.54 | 1.37 |
| 1nc3 | -10.48 | -10.21 | 0.33 | -10.21 | 0.33 | -10.21 | 0.33 | -10.21 | 0.33 | -5.46 | 1.13 |
| 1nvq | -12.99 | -12.46 | 0.72 | -11.68 | 0.75 | -11.05 | 0.97 | -10.61 | 0.86 | -8.79 | 1.27 |
| 1o0h | -5.44 | -4.59 | 0.52 | -4.59 | 0.52 | -4.59 | 0.52 | -4.12 | 0.52 | -2.13 | 1.69 |
| 1o5b | -7.62 | -7.42 | 0.47 | -7.14 | 0.78 | -6.68 | 1.56 | -6.57 | 0.79 | -5.99 | 0.88 |
| 1p1n | -10.49 | -10.36 | 0.37 | -10.36 | 0.37 | -9.17 | 0.56 | -8 | 1.33 | -5.47 | 4.7 |
| 1p1q | -8.34 | -8.74 | 0.93 | -8.65 | 0.8 | -8.65 | 0.8 | -8.16 | 0.49 | -5.99 | 1.54 |
| 1ps3 | -6.12 | -6.36 | 0.83 | -6.36 | 0.83 | -5.51 | 0.71 | -5.51 | 0.71 | -3.97 | 1.36 |
| 1pxn | -6.81 | -8.01 | 0.83 | -8.01 | 0.83 | -7.82 | 0.87 | -7.01 | 6.22 | -6.06 | 1.66 |
| 1q8t | -6.22 | -6.75 | 2.35 | -6.47 | 2.45 | -6.47 | 2.45 | -6.47 | 2.45 | -5.09 | 0.83 |
| 1qf1 | -9.35 | -9.04 | 0.27 | -9.04 | 0.27 | -8.18 | 0.61 | -5.35 | 0.91 | -5.35 | 0.91 |
| 1qkt | -8.71 | -9.05 | 0.95 | -8.44 | 0.85 | -8.25 | 1.08 | -8.19 | 1.47 | -6.83 | 6.93 |
| 1r5y | -6.49 | -5.95 | 0.5 | -5.65 | 0.81 | -4.55 | 1.39 | -4.55 | 1.39 | -3.61 | 6.44 |
| 1s38 | -7.22 | -6.56 | 0.56 | -6.56 | 0.56 | -6.56 | 0.56 | -6.56 | 0.56 | -4.4 | 5.24 |
| 1syi | -13.11 | -13.26 | 0.24 | -13.26 | 0.24 | -11.57 | 0.41 | -10.34 | 0.76 | -0.87 | 4.8 |
| 1uto | -2.93 | -5 | 3.58 | -5 | 3.58 | -4.9 | 1.97 | -4.32 | 2.19 | -3.81 | 2.22 |
| 1vso | -6.4 | -6.78 | 0.68 | -6.77 | 0.42 | -6.77 | 0.42 | -5.03 | 0.97 | -4.27 | 1.7 |
| 1w4o | -5.35 | -4.99 | 0.55 | -4.88 | 0.49 | -4.88 | 0.49 | -4.58 | 0.87 | -3.95 | 1.3 |
| 1y6r | -11.48 | -11.64 | 0.3 | -11.64 | 0.3 | -11.64 | 0.3 | -9.38 | 0.58 | -6.22 | 0.97 |
| 1yc1 | -8.01 | -7.94 | 0.56 | -7.9 | 0.79 | -7.9 | 0.79 | -7.58 | 0.88 | -6.85 | 1.11 |
| 1ydr | -8 | -7.86 | 1.05 | -7.86 | 1.05 | -7.86 | 1.05 | -7.72 | 1.1 | -6.52 | 1.14 |
| 1z95 | -11.42 | -11.31 | 0.54 | -11.31 | 0.54 | -11.31 | 0.54 | -9.7 | 0.74 | -8.74 | 1.32 |
| 1z9g | -5.42 | -5.55 | 0.61 | -5.4 | 0.64 | -5.4 | 0.64 | -5.4 | 0.64 | -3.67 | 1.25 |
| 2al5 | -11.58 | -10.72 | 0.66 | -10.72 | 0.66 | -10.72 | 0.66 | -8.47 | 0.83 | -2.78 | 5.72 |
| 2brb | -7.9 | -7.87 | 0.94 | -7.87 | 0.94 | -7.73 | 0.82 | -7.4 | 0.71 | -7.31 | 0.97 |
| 2c3i | -8.48 | -8.66 | 0.59 | -8.66 | 0.59 | -8.66 | 0.59 | -8.66 | 0.59 | -8.66 | 0.59 |
| 2cbv | -9.5 | -8.95 | 0.55 | -8.9 | 0.6 | -8.9 | 0.6 | -7.24 | 1.21 | -5.57 | 4.09 |
| 2cet | -9.86 | -8.59 | 0.43 | -8.59 | 0.43 | -6.49 | 1.54 | -6.49 | 1.54 | -4.5 | 1.74 |
| 2hb1 | -7.8 | -7.71 | 0.31 | -7.71 | 0.31 | -7.71 | 0.31 | -7.38 | 1.13 | -5.39 | 1.35 |
| 2j78 | -8.09 | -6.63 | 0.55 | -6.63 | 0.55 | -5.13 | 0.99 | -4.22 | 4.27 | -2.99 | 4.72 |
| 2j7h | -8.4 | -9.76 | 0.45 | -9.76 | 0.45 | -8.44 | 0.63 | -8.44 | 0.63 | -7.69 | 2.44 |
| 2pog | -8.31 | -9.02 | 0.49 | -9 | 0.62 | -8.79 | 0.36 | -8.79 | 0.36 | -6.86 | 1.76 |
| 2qbr | -10.27 | -10.26 | 0.6 | -10.26 | 0.6 | -9.25 | 0.65 | -9.25 | 0.65 | -5.62 | 1.93 |
| 2qe4 | -8.68 | -8.93 | 0.43 | -8.35 | 0.62 | -8.35 | 0.62 | -8.35 | 0.62 | -8.35 | 0.62 |
| 2qnq | -18.11 | -16.66 | 0.43 | -16.66 | 0.43 | -16.66 | 0.43 | -15.02 | 0.63 | -2.05 | 0.91 |
| 2r9w | -9.1 | -9.14 | 0.71 | -9.14 | 0.71 | -9.14 | 0.71 | -7.44 | 1.28 | -6.08 | 6.65 |

| | | | | | | | | | | |
|---|---|---|---|---|---|---|---|---|---|---|
| 2v00 | -5.14 | -5.22 | 0.65 | -5.22 | 0.65 | -5.22 | 0.57 | -5.1 | 0.7 | -3.82 | 1.16 |
| 2vvn | -7.84 | -6.99 | 0.54 | -6.69 | 5.84 | -6.69 | 5.84 | -6.69 | 5.84 | -6.09 | 0.95 |
| 2w4x | -6.29 | -6.67 | 0.62 | -6.33 | 0.8 | -5.95 | 0.97 | -5.15 | 0.99 | -3.87 | 1.43 |
| 2weg | -4.39 | -4.56 | 1.77 | -4.56 | 1.77 | -4.51 | 1.11 | -4.34 | 1.16 | -3.92 | 1.11 |
| 2wnc | -8.01 | -7.95 | 0.47 | -7.83 | 2.26 | -7.83 | 2.26 | -7.72 | 2.34 | -6.3 | 1.38 |
| 2wvt | -7.57 | -6.63 | 0.9 | -6.27 | 0.6 | -5.89 | 3.54 | -5.61 | 3.53 | -4.45 | 2.66 |
| 2xb8 | -10.93 | -9.7 | 0.74 | -9.7 | 0.74 | -9.7 | 0.74 | -5.49 | 1.36 | -5.49 | 1.36 |
| 2xdl | -4.45 | -4.43 | 0.61 | -4.31 | 5.56 | -4.31 | 5.56 | -4.02 | 5.57 | -3.92 | 0.9 |
| 2xj7 | -10.04 | -9.38 | 0.33 | -9.38 | 0.33 | -9.38 | 0.33 | -9.38 | 0.33 | -9.38 | 0.33 |
| 2yfe | -7.52 | -7.63 | 0.45 | -6.48 | 0.64 | -6.34 | 1 | -4.84 | 1.66 | 0.93 | 2.07 |
| 2ymd | -8.91 | -8.37 | 0.4 | -8.37 | 0.4 | -8.37 | 0.4 | -7.27 | 0.67 | -3.8 | 5.16 |
| 2zda | -12.94 | -12.61 | 0.62 | -12.61 | 0.62 | -12.61 | 0.62 | -12.51 | 0.55 | -4.28 | 1.48 |
| 3acw | -8.97 | -8.83 | 0.59 | -8.67 | 0.64 | -8.67 | 0.64 | -8.46 | 0.35 | -7.31 | 1.13 |
| 3ao4 | -5.46 | -5.59 | 0.68 | -5.59 | 0.68 | -5.45 | 0.77 | -5.2 | 1.08 | -4.1 | 6.9 |
| 3aru | -4.64 | -4.83 | 6.98 | -4.83 | 6.98 | -4.83 | 6.98 | -4.29 | 6.71 | -4.21 | 2.73 |
| 3arv | -5.79 | -6.66 | 7.79 | -6.6 | 7.69 | -6.53 | 7.07 | -6.53 | 7.31 | -6.26 | 7.64 |
| 3ary | -4.48 | -5.89 | 7.22 | -5.69 | 7.3 | -5.5 | 7.49 | -5.5 | 7.49 | -5.17 | 4.1 |
| 3b27 | -5.13 | -5.18 | 1.24 | -5.18 | 1.15 | -5.01 | 1.19 | -4.85 | 0.4 | -4.85 | 0.4 |
| 3bgz | -10.02 | -10.07 | 0.49 | -9.83 | 0.89 | -9.83 | 0.89 | -9.22 | 0.87 | -8.1 | 1.47 |
| 3cj4 | -7.98 | -9.1 | 0.85 | -8.69 | 0.34 | -8.3 | 1.46 | -8.12 | 1.54 | -7.42 | 1.17 |
| 3d4z | -5.54 | -5.5 | 0.36 | -5.5 | 0.36 | -5.5 | 0.36 | -5.5 | 0.36 | -2.98 | 3.65 |
| 3d6q | -3.36 | -4.52 | 2.63 | -4.51 | 2.27 | -4.51 | 2.27 | -4.51 | 2.27 | -3.66 | 2.75 |
| 3dd0 | -5.7 | -6.18 | 0.65 | -6.18 | 0.65 | -6.18 | 0.65 | -5.14 | 1.55 | -5.14 | 1.55 |
| 3dx1 | -8.15 | -8.72 | 2.33 | -8.72 | 2.33 | -8.72 | 2.33 | -8.56 | 2.38 | -5.15 | 3.76 |
| 3dx2 | -10.32 | -9.42 | 0.78 | -9.42 | 0.78 | -9.42 | 0.78 | -8.59 | 1 | -7.59 | 2.8 |
| 3dxg | -2.42 | -3.92 | 6.57 | -3.4 | 5.5 | -3.31 | 5.59 | -3.31 | 5.59 | -2.45 | 5.3 |
| 3ebp | -8.47 | -8.98 | 0.92 | -8.98 | 0.92 | -8.98 | 0.92 | -8.49 | 5.78 | -8.49 | 5.78 |
| 3ehy | -4.6 | -6.04 | 3.69 | -5.93 | 3.69 | -5.93 | 3.69 | -5.65 | 3.66 | -5.22 | 3.69 |
| 3f3a | -7.9 | -7.46 | 0.86 | -7.46 | 1.11 | -7.46 | 1.11 | -6.53 | 1.47 | -6.53 | 1.47 |
| 3f3c | -8.87 | -8.82 | 0.31 | -8.82 | 0.31 | -6.19 | 0.57 | -5.35 | 2.29 | -3.42 | 2.88 |
| 3f3d | -7.07 | -6.3 | 0.61 | -6.3 | 0.61 | -5.79 | 0.58 | -4.15 | 1.48 | -3 | 2.55 |
| 3f3e | -7.62 | -7.32 | 0.56 | -7.32 | 0.56 | -6 | 0.82 | -5.2 | 1.26 | -5.2 | 1.26 |
| 3fcq | -1.41 | -3.01 | 3.78 | -3.01 | 3.78 | -3.01 | 3.78 | -3.01 | 3.78 | -2.75 | 4.07 |
| 3fur | -13.48 | -13.12 | 0.68 | -13.12 | 0.68 | -12.63 | 0.5 | -11.75 | 0.82 | -11.75 | 0.82 |
| 3fv1 | -16.15 | -14.78 | 0.79 | -14.78 | 0.79 | -14.78 | 0.79 | -14.78 | 0.79 | -14.78 | 0.79 |
| 3fv2 | -15.94 | -14.05 | 0.63 | -13.23 | 0.46 | -13.23 | 0.46 | -11.66 | 0.7 | -9.22 | 0.79 |
| 3g0w | -9.52 | -9.4 | 0.33 | -9.4 | 0.33 | -9.4 | 0.33 | -8.68 | 0.61 | -4.28 | 1.47 |
| 3g2n | -8.32 | -7.66 | 0.59 | -6.78 | 0.75 | -6.78 | 0.75 | -5.47 | 7.18 | -3.42 | 7.48 |
| 3g2z | -3.65 | -5.11 | 5.03 | -5.11 | 5.03 | -4.91 | 5.09 | -4.81 | 5.04 | -4.58 | 4.64 |
| 3g31 | -3.43 | -5.41 | 3.14 | -5.41 | 3.14 | -5.29 | 2.71 | -5.29 | 2.71 | -4.58 | 2.67 |
| 3gbb | -12.28 | -10.76 | 0.67 | -10.45 | 0.73 | -10.26 | 0.63 | -10.15 | 0.91 | -5.39 | 4.55 |
| 3gc5 | -9.98 | -7.72 | 0.59 | -7.72 | 0.59 | -7.72 | 0.59 | -5.22 | 1.83 | -4.66 | 1.17 |
| 3ge7 | -13.87 | -11.9 | 0.98 | -11.9 | 0.98 | -11.68 | 0.56 | -9.37 | 0.83 | -8.33 | 0.98 |
| 3gr2 | 3.2 | -5.89 | 3.83 | -5.89 | 3.83 | -5.89 | 3.83 | -5.89 | 3.83 | -4.83 | 5.66 |
| 3gv9 | -5.17 | -5.07 | 0.92 | -5.07 | 0.92 | -5.07 | 0.92 | -5.07 | 0.92 | -4.69 | 0.48 |
| 3gy4 | -5.76 | -5.28 | 2.51 | -5.28 | 2.51 | -5.21 | 0.77 | -4.87 | 2.35 | -4.71 | 2.17 |
| 3jvr | -7.56 | -7.67 | 0.55 | -7.67 | 0.55 | -7.26 | 0.67 | -6.77 | 0.91 | -6.77 | 0.91 |
| 3jvs | -9.33 | -9.62 | 0.56 | -8.86 | 0.37 | -7.77 | 0.61 | -7.66 | 1.15 | -5.96 | 1.03 |
| 3jya | -6.35 | -7.22 | 1.03 | -7.22 | 1.03 | -7.22 | 1.03 | -6.76 | 0.97 | -6.32 | 5.44 |
| 3kgp | -3.5 | -5.4 | 1.77 | -5.29 | 2.49 | -5.29 | 2.49 | -5.29 | 2.49 | -4.07 | 1.4 |
| 3kr8 | -10.38 | -10.07 | 0.83 | -10.07 | 0.83 | -9.6 | 0.61 | -9.6 | 0.61 | -5.81 | 7.22 |
| 3kwa | 17.11 | 15.46 | 2.42 | 15.46 | 2.42 | 15.46 | 2.42 | 16.49 | 2.69 | 16.49 | 2.69 |
| 3l7b | -5.67 | -5.12 | 0.69 | -5.12 | 0.69 | -4.54 | 0.73 | -4.41 | 0.95 | -3.1 | 1.84 |

| | | | | | | | | | | |
|---|---|---|---|---|---|---|---|---|---|---|
| 3lka | -5.37 | -5.58 | 5.99 | -5.58 | 5.99 | -5.58 | 5.99 | -5.18 | 5.79 | -4.91 | 5.93 |
| 3n76 | -10.11 | -9 | 0.91 | -9 | 0.91 | -7.91 | 1.02 | -5.69 | 0.74 | -4.6 | 1.58 |
| 3n7a | -6.8 | -6.11 | 0.58 | -6.11 | 0.45 | -5.74 | 0.63 | -5.74 | 0.63 | -2.93 | 2.7 |
| 3nq9 | -1.7 | -2.33 | 1.55 | -2.33 | 1.55 | -2.29 | 1.58 | -2.22 | 2.22 | -2.22 | 2.22 |
| 3nx7 | -6.1 | -6.34 | 0.56 | -6.34 | 0.56 | -6.13 | 1.04 | -6.13 | 1.04 | -4.26 | 1.28 |
| 3p5o | -8.4 | -8.84 | 0.65 | -8.82 | 0.58 | -8.82 | 0.58 | -7.9 | 0.97 | -7.2 | 1.12 |
| 3pxf | -7.4 | -8 | 4.69 | -8 | 4.69 | -8 | 4.69 | -7.53 | 6.13 | -6.52 | 4.64 |
| 3pyy | -8.9 | -9.05 | 0.41 | -9.05 | 0.41 | -9.05 | 0.41 | -9.05 | 0.41 | -6.14 | 0.94 |
| 3qqs | -9.15 | -9.18 | 0.51 | -8.67 | 0.97 | -8.48 | 1.49 | -7.95 | 5.1 | -7.42 | 5.06 |
| 3r88 | -5.09 | -5.06 | 0.51 | -4.94 | 0.86 | -4.92 | 0.8 | -4.92 | 0.8 | -3.51 | 5.28 |
| 3rlr | -8.82 | -8.8 | 0.64 | -8.59 | 0.51 | -8.59 | 0.51 | -8.24 | 0.86 | -6.93 | 1.6 |
| 3rr4 | -7.51 | -7 | 0.63 | -6.18 | 0.55 | -6.18 | 0.55 | -5.17 | 1.28 | -4.42 | 1.08 |
| 3rsx | -6.17 | -5.93 | 6.7 | -5.93 | 0.55 | -5.79 | 0.46 | -5.79 | 0.46 | -4.49 | 4.34 |
| 3ryj | -4.88 | -5.15 | 0.77 | -5.09 | 0.58 | -5.09 | 0.8 | -5.09 | 0.8 | -4.68 | 1.03 |
| 3syr | -8.93 | -9.01 | 0.25 | -9.01 | 0.25 | -7.47 | 0.45 | -4.8 | 1.49 | -3.07 | 1.19 |
| 3twp | -2.93 | -3.87 | 2.83 | -3.87 | 2.83 | -3.87 | 2.83 | -3.77 | 2.81 | -2.88 | 2.64 |
| 3u5j | -7.48 | -7.83 | 4.89 | -7.67 | 0.98 | -7.65 | 1.15 | -7.45 | 2.63 | -7.06 | 0.73 |
| 3u9q | -4.54 | -4.76 | 0.81 | -4.76 | 0.81 | -4.76 | 0.81 | -4.3 | 1.33 | -4.3 | 1.33 |
| 3udh | -8.61 | -8.1 | 0.58 | -7.92 | 0.47 | -6.97 | 0.98 | -6.97 | 0.98 | -6.43 | 0.98 |
| 3ui7 | -7.42 | -7.38 | 0.52 | -7.38 | 0.52 | -7.13 | 0.82 | -6.88 | 0.75 | -6.59 | 0.83 |
| 3uo4 | -12.11 | -11.9 | 0.36 | -11.87 | 0.67 | -11.87 | 0.67 | -11.87 | 0.67 | -6.53 | 1.66 |
| 3up2 | -10.04 | -10.21 | 0.44 | -9.99 | 0.54 | -9.63 | 0.56 | -9.63 | 0.56 | -9.63 | 0.56 |
| 3uuo | -7.27 | -7.34 | 0.62 | -7.34 | 0.62 | -7.34 | 0.62 | -7.34 | 0.62 | -6.09 | 1.36 |
| 3wtj | -8.6 | -8.58 | 1.16 | -8.44 | 0.36 | -8.44 | 0.36 | -8.44 | 0.36 | -5.88 | 6.47 |
| 3zdg | -5.6 | -5.56 | 0.29 | -5.37 | 0.89 | -5.17 | 1.06 | -4.78 | 1.04 | -1.78 | 3.01 |
| 3zsx | -10.11 | -9.83 | 0.28 | -9.79 | 0.92 | -9.79 | 0.92 | -9.52 | 0.45 | -7.47 | 1.67 |
| 3zt2 | -8.31 | -8.33 | 0.58 | -8.33 | 0.58 | -8.33 | 0.58 | -7.87 | 0.9 | -7.11 | 0.53 |
| 4abg | -4.98 | -5.12 | 0.7 | -5.09 | 0.77 | -5.09 | 0.77 | -4.79 | 0.71 | -3.79 | 1.34 |
| 4bkt | -6.04 | -6.07 | 0.44 | -6.07 | 0.44 | -6.07 | 0.44 | -6.07 | 0.44 | -5.21 | 0.43 |
| 4cig | -10.7 | -10.79 | 0.92 | -10.79 | 0.92 | -9.97 | 0.42 | -9.97 | 0.42 | -6.72 | 1.81 |
| 4ciw | -6.27 | -6.88 | 0.72 | -6.88 | 0.72 | -6.46 | 0.74 | -5.91 | 0.73 | -3.29 | 4.25 |
| 4cr9 | -4.35 | -5.87 | 6.13 | -5.53 | 6.23 | -5.25 | 5.88 | -5.03 | 5.78 | -4.85 | 5.88 |
| 4ddh | -6.72 | -7.03 | 4.69 | -6.99 | 4.66 | -6.99 | 4.66 | -6.27 | 0.7 | -5.3 | 1.64 |
| 4ddk | -4.19 | -4.5 | 4.23 | -4.5 | 4.94 | -4.5 | 4.23 | -4.35 | 4.36 | -3.96 | 4.89 |
| 4de1 | -7.56 | -7.39 | 0.64 | -7.25 | 0.43 | -7.13 | 0.53 | -4.55 | 2.16 | -4.54 | 1.17 |
| 4djv | -11.71 | -10.57 | 0.66 | -10.51 | 0.54 | -10.51 | 0.54 | -10.2 | 0.89 | -6.53 | 7.38 |
| 4dld | -11.42 | -10.87 | 0.29 | -10.86 | 0.35 | -10.05 | 0.49 | -7.63 | 1.23 | -5.13 | 5.34 |
| 4dli | -8.57 | -8.79 | 1 | -8.79 | 1 | -8.6 | 0.57 | -8.52 | 1.11 | -6.76 | 1.45 |
| 4e6q | -8.44 | -8.35 | 0.51 | -8.3 | 0.65 | -8.3 | 0.65 | -7.53 | 1.09 | 0.05 | 1.08 |
| 4eo8 | -8.78 | -9.08 | 0.74 | -9.08 | 0.7 | -9.05 | 0.6 | -8.73 | 0.82 | -6.83 | 0.94 |
| 4eor | -10.55 | -10.39 | 0.48 | -10.31 | 0.72 | -10.31 | 0.72 | -10.31 | 0.72 | -8.35 | 0.96 |
| 4f09 | -7.2 | -7.4 | 1.1 | -7.4 | 1.1 | -6.77 | 3.67 | -6.17 | 3.69 | -6.07 | 1.63 |
| 4f2w | -10.69 | -10.06 | 0.45 | -9.81 | 0.56 | -9.75 | 0.55 | -8.94 | 0.78 | -8.24 | 0.98 |
| 4f9w | -9.92 | -10.2 | 0.6 | -9.97 | 0.99 | -9.97 | 0.99 | -9.35 | 0.81 | -8.66 | 0.74 |
| 4gfm | -7.43 | -7.38 | 0.43 | -7.38 | 0.43 | -7.38 | 0.43 | -7.38 | 0.43 | -6.22 | 6.2 |
| 4gkm | -9.22 | -10.5 | 5.91 | -10.5 | 5.91 | -10.37 | 5.87 | -10.37 | 5.87 | -10.02 | 5.63 |
| 4hge | -9.73 | -9.44 | 0.94 | -9.44 | 0.94 | -9.42 | 0.68 | -8.86 | 0.91 | -8.46 | 5.31 |
| 4ih5 | 8.76 | -4.89 | 0.88 | -4.88 | 1.46 | -4.88 | 1.46 | -4.09 | 1.07 | -3.79 | 2.33 |
| 4ih7 | -7.13 | -7.22 | 0.52 | -7.22 | 0.52 | -7.16 | 0.62 | -6.56 | 0.72 | -5.87 | 5.52 |
| 4ivb | -9.44 | -9.6 | 0.22 | -9.49 | 0.47 | -9.49 | 0.47 | -6.73 | 1.56 | -5.52 | 4.35 |
| 4ivc | -9.71 | -9.51 | 0.49 | -9.32 | 0.55 | -9.28 | 0.68 | -8.95 | 0.57 | -6.24 | 0.8 |
| 4ivd | -9.98 | -9.53 | 0.77 | -9.53 | 0.77 | -9.53 | 0.77 | -9.36 | 0.84 | -5.77 | 1.26 |
| 4j21 | -10.24 | -10.13 | 0.84 | -9.64 | 1.04 | -9.09 | 0.82 | -8.83 | 0.82 | -8.31 | 1.16 |

| | | | | | | | | | | |
|---|---|---|---|---|---|---|---|---|---|---|
| 4j28 | -7.51 | -6.83 | 0.51 | -6.82 | 0.64 | -6.59 | 0.77 | -6.59 | 0.77 | -6.54 | 2.01 |
| 4jfs | -5.77 | -6.75 | 0.55 | -6.75 | 0.55 | -6.75 | 0.55 | -6.2 | 0.61 | -5.95 | 0.66 |
| 4jsz | -3.06 | -4.72 | 2.81 | -4.55 | 4.09 | -4.25 | 3.01 | -4.23 | 4.21 | -4.23 | 4.21 |
| 4jxs | -9.05 | -8.64 | 0.45 | -8.64 | 0.45 | -7.37 | 0.77 | -7.37 | 0.77 | -5.54 | 1.66 |
| 4k18 | -9.76 | -10.03 | 0.6 | -10.03 | 0.6 | -10.03 | 0.6 | -9.83 | 0.57 | -6.51 | 6.41 |
| 4k77 | -6.94 | -6.66 | 0.47 | -6.66 | 0.47 | -6.66 | 0.47 | -6.26 | 0.75 | -5.91 | 2.86 |
| 4kz6 | -4.43 | -5.05 | 1.35 | -5.05 | 1.35 | -5.05 | 1.35 | -5.05 | 1.35 | -3.79 | 5.44 |
| 4kzq | -8.6 | -8.5 | 0.45 | -8.3 | 0.88 | -8.06 | 0.74 | -7.87 | 0.96 | -6.86 | 1.14 |
| 4kzu | -10.48 | -10.15 | 0.53 | -9.5 | 0.75 | -9.2 | 1.04 | -7.4 | 1.04 | -6.93 | 1.08 |
| 4llx | -3.86 | -3.94 | 2.79 | -3.92 | 2.75 | -3.92 | 2.75 | -3.92 | 2.75 | -3.61 | 2.72 |
| 4lzs | -5.96 | -6.14 | 1.11 | -6.14 | 1.11 | -6.05 | 0.65 | -5.63 | 0.76 | -5.62 | 0.68 |
| 4m0y | -7.26 | -7.3 | 7.09 | -7.3 | 7.09 | -7.13 | 7.08 | -6.97 | 1.49 | -6.75 | 5.16 |
| 4m0z | -13.44 | -12.65 | 0.44 | -12.65 | 0.44 | -12.22 | 0.63 | -11.61 | 0.59 | -6.49 | 4.59 |
| 4mgd | -4.28 | -7.96 | 0.58 | -7.9 | 0.39 | -7.9 | 0.39 | -7.9 | 0.39 | -7.21 | 1.16 |
| 4mme | -9.79 | -9.65 | 0.4 | -9.65 | 0.4 | -9.65 | 0.4 | -9.27 | 0.24 | -8.43 | 0.7 |
| 4ogj | -9.6 | -9.74 | 0.57 | -9.74 | 0.57 | -9.47 | 0.46 | -9.47 | 0.46 | -6.2 | 5.78 |
| 4owm | -4.26 | -5.01 | 3.07 | -4.92 | 0.9 | -4.92 | 0.9 | -4.92 | 0.9 | -4.92 | 0.9 |
| 4pcs | -7.26 | -7.17 | 0.39 | -7.17 | 0.39 | -7.1 | 0.9 | -6.81 | 0.73 | -6.22 | 1.08 |
| 4qac | -10.65 | -10.64 | 0.42 | -10.64 | 0.67 | -10.64 | 0.67 | -10.3 | 0.68 | -8.61 | 1.5 |
| 4u4s | -5.77 | -5.9 | 0.61 | -5.62 | 0.44 | -5.6 | 2.61 | -5.6 | 0.86 | -5.6 | 0.86 |
| 4wiv | -7.29 | -7.64 | 1.02 | -7.62 | 0.52 | -7.54 | 0.57 | -7.54 | 0.57 | -7.15 | 0.96 |
| 5c28 | -4.9 | -5.18 | 3.5 | -5.18 | 3.5 | -5.17 | 3.5 | -4.58 | 3.32 | -3.96 | 4.53 |
| 5dwr | -13.38 | -13.13 | 0.67 | -13.13 | 0.67 | -12.26 | 1 | -10.99 | 1.29 | -10.99 | 1.29 |